\documentstyle[pre,twocolumn,aps,psfig]{revtex}
\begin{document}
\preprint{TAUP 2463-97}
\draft
\twocolumn[\hsize\textwidth\columnwidth\hsize\csname@twocolumnfalse%
\endcsname
\title{Chaos of the Relativistic Parametrically Forced van der Pol Oscillator}
\author{Y. Ashkenazy$^{\text{1}}$\cite{addressA}, C. Goren$^{\text{1}}$
and L. P. Horwitz$^{\text{1,2}}$\cite{addressB}}
\address{$^1$ Department of Physics, Bar-Ilan University, Ramat-Gan 52900,
Israel\\
$^2$ School of Physics, Raymond and Beverly Sackler Faculty of Exact Sciences\\
Tel-Aviv University, Ramat-Aviv, Israel.}
\date{\today}
\maketitle
\begin{abstract}
{
\baselineskip=2.5ex
A manifestly relativistically covariant form of the van der Pol oscillator
in $1+1$ dimensions is studied. We show that the driven relativistic equations,
for which $x$ and $t$ are coupled, relax very quickly to a pair of identical
decoupled equations, due to a rapid vanishing of the ``angular momentum'' 
(the boost in $1+1$ dimensions). A similar effect occurs in the damped driven
covariant Duffing oscillator previously treated. This effect is an example 
of entrainment, or synchronization (phase locking), of coupled chaotic systems.
The Lyapunov exponents are calculated using the very efficient method of Habib
and Ryne. We show a Poincar\'e map that demonstrates this effect and 
maintains remarkable stability in spite of the inevitable accumulation of 
computer error in the chaotic region. For our choice of parameters, the 
positive Lyapunov exponent is about 0.242 almost independently of the 
integration method.}
\end{abstract}
]
\narrowtext
\newpage

The chaotic behavior of a covariant relativistic generalization of the 
classical Duffing oscillator in $1+1$ dimensions has been studied
recently \cite{{Horwitz91},{Horwitz92}}.  This system is completely
integrable, since both the total ``mass'' (the value of the invariant 
generator of dynamical evolution) and the hyperbolic ``angular momentum''
 (the generator of Lorentz transformations in $1+1$ dimensions) are conserved.
Under non-stationary perturbation, including a friction term as well,  
however, the stable and unstable orbits in the phase space separate and cross
(infinitely many times).\footnote{One can think of the friction term as
 the result
of radiation reaction for a charged oscillator, i.e., the damping of a 
radiating dipole which depends only on the relative motion of the charges
\cite{Horwitz92}.}

  Some general features of the resulting chaotic 
orbits were discussed in ref.\cite{Horwitz91}, such as the phenomena of 
period doubling and the apparent existence of a strange attractor 
\cite{Guckenheimer}.
In ref.\cite{Horwitz92}, analytic solutions were studied for the unperturbed
problem in the neighborhood of the separatrix, and it was shown that the
perturbative Melnikov criterion\cite{Melnikov63} for homoclinic chaos was 
satisfied.   It was found that along with chaotic
behavior in space, there is chaos in time in such a system, with perhaps 
profound implications for our perception of the nature of dynamical processes.

In this paper, we study a manifestly covariant dynamical system
with no Hamiltonian structure, i.e., a covariant form of the 
van der Pol oscillator. The model was
originally introduced by van der Pol\cite{vanderPol26} to describe the 
triode electronic
oscillator, and used by van der Pol and van der Mark as a model for
heart rhythms\cite{vanderPol28}. The van der Pol equation was introduced 
before chaos was
 systematically defined or studied, and it became one of the first examples
to be analyzed for chaotic behavior and associated entrainment phenomena
\cite{{Cartwright},{Jackson},{KBM},{Sanders}}.
 This system carries intrinsic (coordinate dependent)
 friction; if this term vanishes, one obtains the structure of 
an ordinary harmonic oscillator. The method of Melnikov is not applicable for 
the van der Pol system, since there is no homoclinic orbit associated with 
the unperturbed problem. We shall therefore calculate the Lyapunov exponents 
to demonstrate the chaotic behavior \cite{Melnikov63}.

It was observed in the case of the relativistic Duffing problem 
that the $x$ and
$t$ modes, viewed as two coupled systems, could converge, in the 
presence of friction, to a single Duffing type oscillator (the ``angular
momentum'' converges to zero).  This convergence is an example of the entrainment (sometimes
called ``phase locking'') of two coupled chaotic systems.  We observe a 
similar behavior in the relativistic van der Pol equations.
It is conceivable that nonconservative relativistic electronic
systems have similar properties, providing examples of fundamental interest 
for the
entrainment, or synchronization (phase locking) properties of such chaotic 
systems \cite{Hilborn94}.

As for the Duffing oscillator, we add a driving force that conserves
the ``angular momentum'', so that there is minimal distortion introduced by
such a term. We find, as for the Duffing oscillator, that the time and 
space modes coalesce, with resulting loss of independent degrees of freedom.
 The resulting
one-dimensional system does not coincide precisely with the usual 
nonrelativistic
case; the forcing term appears proportional to the dependent variable
 ($x(\tau)$ or $t(\tau)$).  This term must be then shifted to the 
non-linear coefficient
of the dependent variable in the van der Pol equation (previous studies
 have considered driving terms in the 
coefficient of the $\dot x$\cite{Shaw81}). It was therefore necessary to
investigate the chaotic 
structure of the resulting system independently of 
previous studies. 

Although the chaotic behavior of the usual van der Pol
system is difficult to observe, we found that in our case, the usual methods
utilizing, for example, Poincar\'e maps, clearly demonstrate the more complex 
chaotic character of this system. Moreover, we employ a method
given by Habib and Ryne \cite{Habib} for the study of the Lyapunov 
exponents to confirm the chaotic nature of the system; this method is very
effective in studying a system of this type, i.e., a system for which the 
local linearization can be described in terms of a Hamiltonian form.

The rapid relaxation of the dissipative relativistic system to a system
closely related to the nonrelativistic form is of interest in itself in 
providing an example of how dissipation can induce an effectively 
nonrelativistic dynamics (through entrainment of the time and space modes).
  A similar phenomena occurs, as mentioned above, in the dissipative form
of the Duffing oscillator under certain conditions \cite{Horwitz91}.

Following the classical relativistic mechanics of Stueckelberg 
\cite{Stueckelberg41}, and its extension \cite{Horwitz73}
to the many body problem,
one can define a relativistic invariant evolution function $K$ (analogous to 
the nonrelativistic Hamiltonian), which generates ``Hamilton-like''
equations:
\begin{equation} \label{e1}
\frac{dx_i^{\mu}}{d\tau} = \frac{\partial K}{\partial p_{i\mu}}, \qquad
\frac{dp_i^{\mu}}{d\tau} = -\frac{\partial K}{\partial x_{i\mu}}, 
\end{equation}
where $p_i^{\mu}$ and $x_i^{\mu}$ are the energy-momentum and spacetime 
coordinates ($i=1\dots N$, $\mu = 0,1$, with metric -, +) and $\tau$ is a 
universal invariant parameter for motion in $2N$-dimensional phase space.

Consider the two body problem with a potential that depends 
on the invariant distance between them, for which the
 corresponding evolution function, $K$,
may be written as:
\begin{equation} \label{e2}
K = \frac{p_1^\mu p_{1\mu}}{2M_1} + \frac{p_2^\mu p_{2\mu}}{2M_2} +V(\rho^2),
\end{equation}
where $\rho^2$ is the Poincar\'e invariant
\begin{eqnarray} \label{e3}
\rho^2 &=& (x_1^\mu-x_2^\mu)(x_{1\mu}-x_{2\mu}) \nonumber \\
&=& (x_1-x_2)^2-(t_1-t_2)^2.
\end{eqnarray}
It is possible to write the two body problem as a function of center of mass
and the relative motion between the bodies (in a way similar to nonrelativistic
mechanics):
\begin{eqnarray} \label{e4}
&&K = K_{c.m.}+K_{rel} \nonumber \\
&&K_{c.m.} = \frac{P^\mu P_\mu}{2M} \\
&&K_{rel} = \frac{p^\mu p_\mu}{2m}+V(\rho^2),\nonumber
\end{eqnarray}
where
\begin{eqnarray}
&&M=M_1+M_2 \qquad m=\frac{M_1 M_2}{M_1+M_2} \nonumber \\
&&P^\mu = p_1^\mu + p_2^\mu \qquad 
p^\mu = \frac{M_2 p_1^\mu - M_1 p_2^\mu}{M_1+M_2} \nonumber \\
&&X^\mu = \frac{M_1 x_1^\mu + M_2 x_2^\mu}{M_1+M_2} \qquad
 x^\mu=x_1^\mu-x_2^\mu. \nonumber
\end{eqnarray}
\begin{figure}[thb]
\psfig{figure=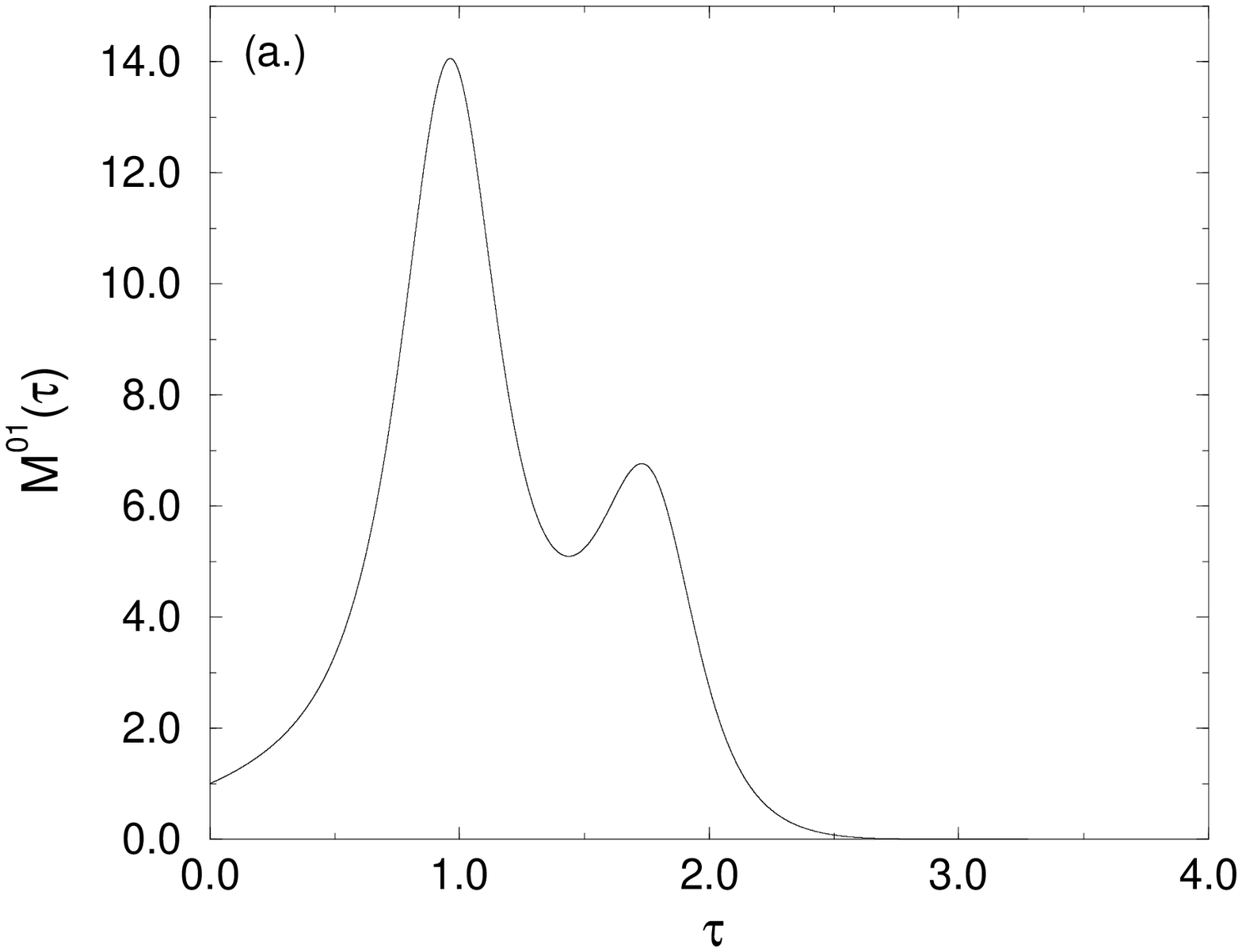,height=8cm,width=8.5cm}
\psfig{figure=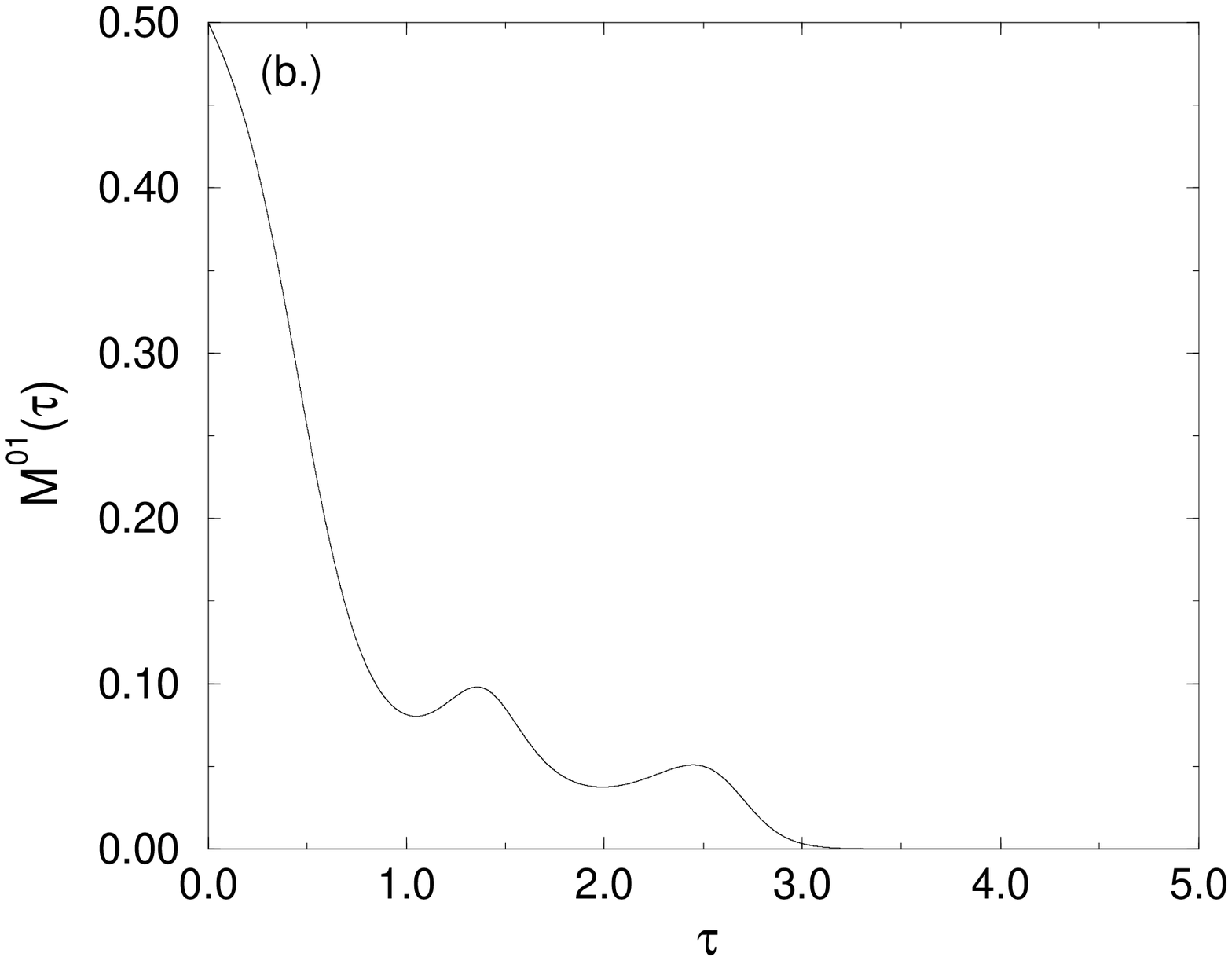,height=8cm,width=8.5cm}
\caption{\label{fig1}
The angular momentum, $M^{01}(\tau)$, of the typical cases where 
there is a) exponential growth followed by an oscillatory exponential decay of 
$M^{01}(\tau)$ where $\dot M_0^{01} M_0^{01} > 0$, 
b) oscillatory exponential decay of $M^{01}(\tau)$ where 
$\dot M_0^{01} M_0^{01} \le 0$
}
\end{figure}
The equations of motion in the new variables are  
\begin{eqnarray} \label{e5}
\frac{d x^\mu}{d \tau} = \frac{\partial K_{rel}}{\partial p_\mu} \qquad
\frac{d p^\mu}{d \tau} = -\frac{\partial K_{rel}}{\partial x_\mu} \\
\frac{d X^\mu}{d \tau} = \frac{\partial K_{c.m.}}{\partial P_\mu} \qquad
\frac{d P^\mu}{d \tau} = 0; \nonumber
\end{eqnarray}
the system separates to two equations describing the center of mass and the 
relative motion of the system.

\begin{figure}[t]
\psfig{figure=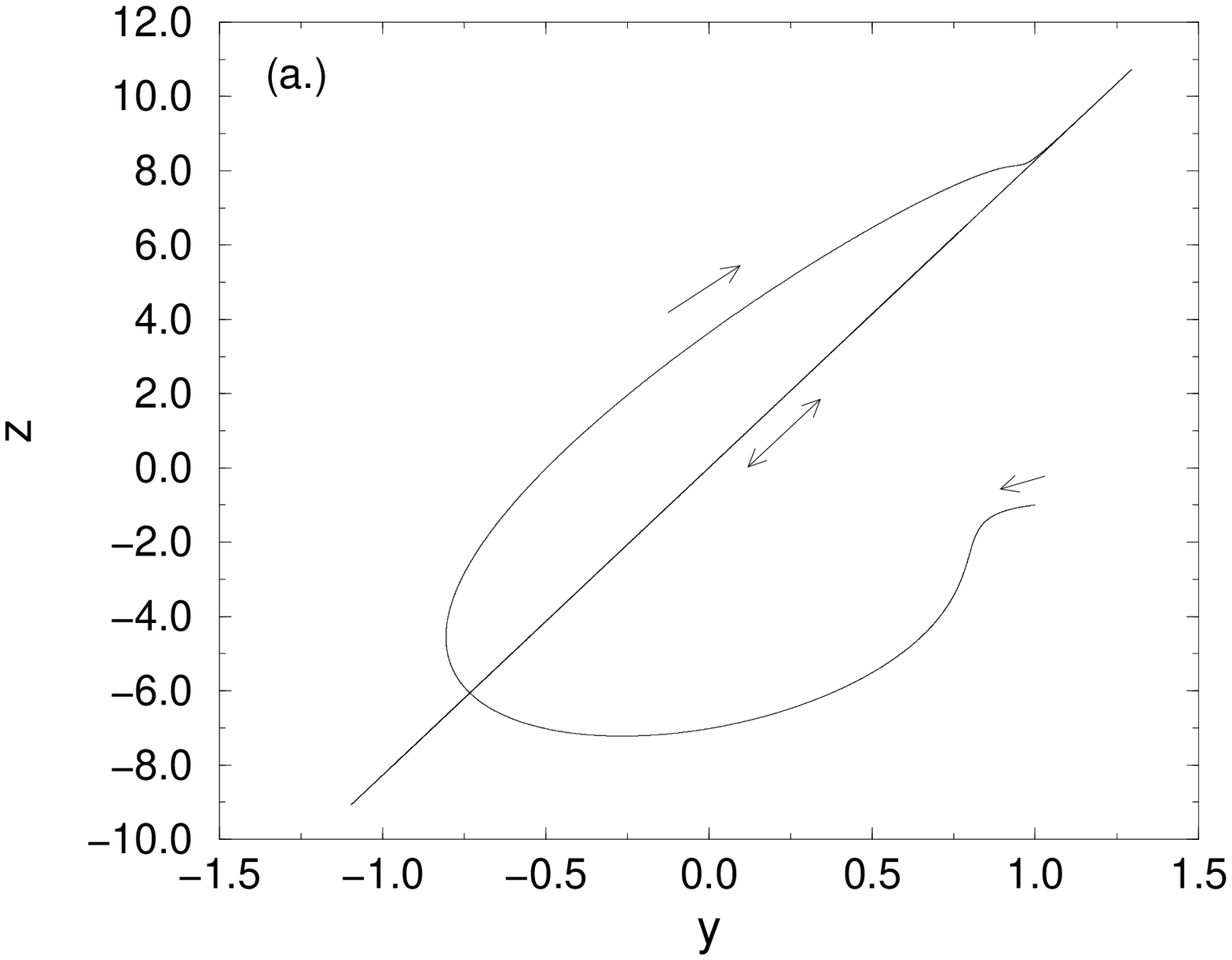,height=8cm,width=8.5cm}
\psfig{figure=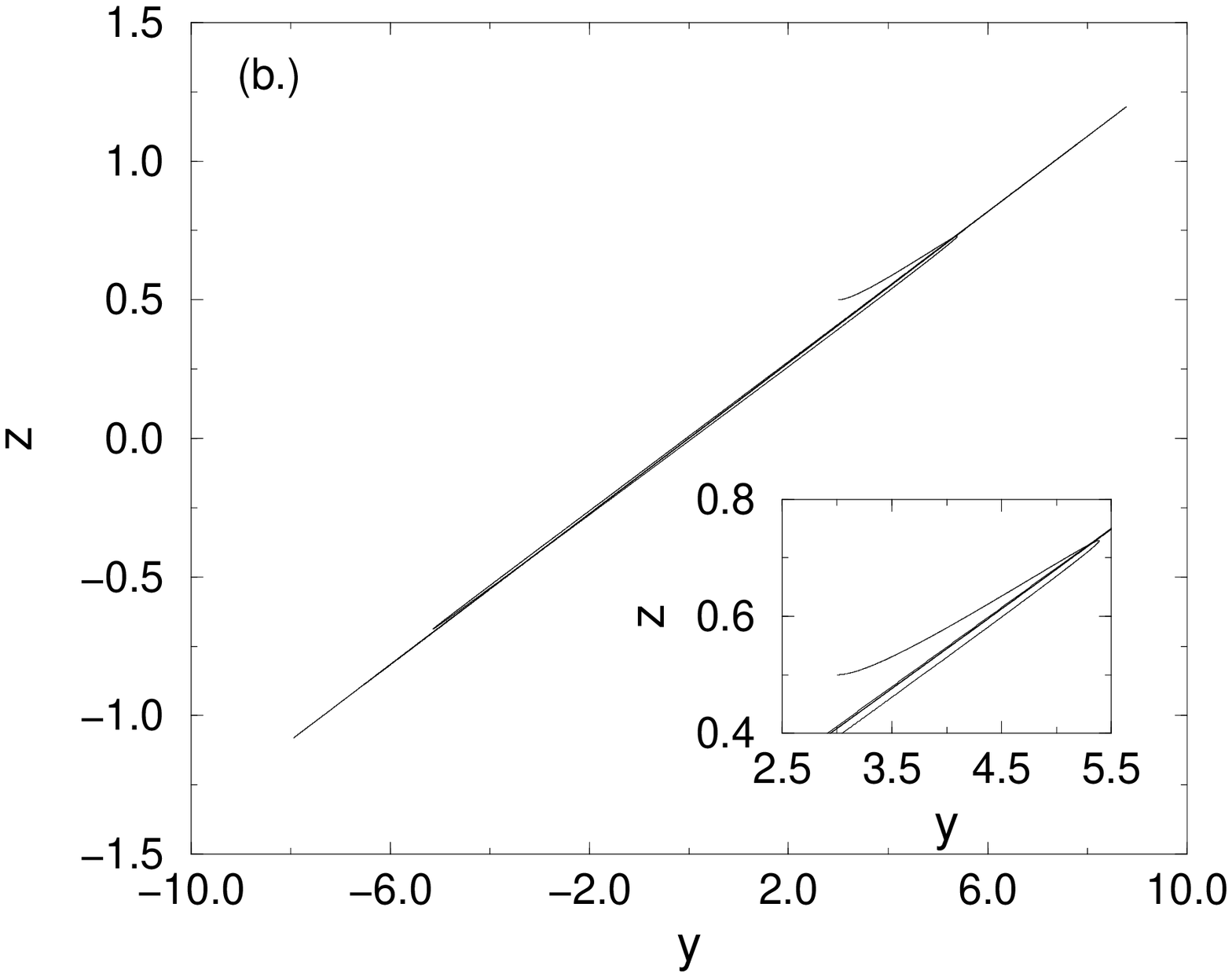,height=8cm,width=8.5cm}
\caption[tbp]{\label{fig2}
The $(y,z)$ plane of fig. \ref{fig1}a and b. 
The inset in fig. \ref{fig2}b shows an enlargement of the turning point from 
the initial point to the final linear curve.
}
\end{figure}
The equation for the nonrelativistic externally forced van der Pol 
oscillator is \cite{{vanderPol26},{vanderPol28}} 
\footnote{Jackson \cite{Jackson} has shown that a self-stabilizing radiating 
system can be described by this equation, where $x$ corresponds to $dx'/dt$ of
the original system. The quantity ${dx^\mu}'/d\tau$ occurs in the relativistic 
form of such a system, where $d\tau$ is an invariant.  The same substitution
would then be valid in this case.} :
\begin{equation} \label{e6}
\ddot x + \alpha (x^2-1)\dot x +kx = g\cos \omega t,
\end{equation}
where the right hand side corresponds to an external driving force.
Although there is no Hamiltonian which generates this equation, we may 
nevertheless consider its relativistic generalization in terms of a 
relative motion problem, for which there is no evolution function $K$.

In this paper, we study the relativistic generalization of Eq. (\ref{e6}) 
in the form
\begin{equation} \label{e7}
\ddot x^\mu + \alpha (\rho^2 -1) \dot x^\mu +x^\mu =0, 
\end{equation}
or, in terms of components, 
\begin{eqnarray} \label{e8}
&\ddot x + \alpha(x^2 -t^2 -1) \dot x +kx = &0\\
&\ddot t + \alpha(x^2 -t^2 -1) \dot t +kt = &0,
\end{eqnarray}
where $x,t$ are the relative coordinates of the two body system.  The term
$\dot x$ can be understood as representative of  friction due to radiation, as 
for damping due to dipole radiation of two charged particles, proportional to
the relative velocity $\dot x^\mu$ in the system \cite{Horwitz92}.  To study 
the existence of chaotic behavior on this system, we add a driving 
force\footnote{We understand $\ddot x^\mu$ to be proportional to force, 
from its
form in the framework of the Hamiltonian theories.}
to the system (it already contains dissipation intrinsically) in such a 
way that it does not provide a mechanism in addition to the dissipative
terms for the change of ``angular momentum'' (this quantity is conserved by 
eq. (\ref{e7}) with $\alpha=0$)
\begin{equation}\label{e9}
M^{01} = x^0 p^1 - x^1 p^0 = m (t\dot x - x \dot t).
\end{equation}
To achieve this, we take a driving force proportional to $x^\mu$,
so that (\ref{e8}) becomes
\begin{eqnarray}
\ddot x + \alpha(x^2 -t^2 -1) \dot x +kx &= &fx\cos \omega \tau \label{e10a}\\
\ddot t + \alpha(x^2 -t^2 -1) \dot t +kt &= &ft\cos \omega \tau.\label{e10b}
\end{eqnarray}
 The derivative of the angular momentum can be evaluated in terms of these
equations as
\begin{equation} \label{e11}
\frac{dM^{01}}{d \tau} = - \alpha (x^2-t^2-1) M^{01}.
\end{equation}
In contrast to exponential decay of the angular momentum that was shown in 
Duffing oscillator \cite{Horwitz91}, one cannot see clearly whether eq. 
(\ref{e11}) reflects such decay or not. Exponential decay of $M^{01}$ means 
that after a relatively short time $M^{01}$ goes to zero, and that 
$x$ becomes proportional to $t$ ($x = \beta t$). Thus, in this case, 
the two coupled
equations would actually reduce to one equation. 
In order to simplify the investigation of eqs. (\ref{e10a}) and (\ref{e10b}),
one can introduce two symmetric equations by subtracting and adding them, to
obtain 
\begin{eqnarray}
\ddot y + \alpha(yz -1) \dot y +ky &= &fy\cos \omega \tau, \label{e12a}\\
\ddot z + \alpha(yz -1) \dot z +kz &= &fz\cos \omega \tau,\label{e12b}
\end{eqnarray}
where $y=x-t$ and $z=x+t$. Eqs. (\ref{e12a}), (\ref{e12b}) are completely
symmetric and it is clear that if one starts from initial conditions
$(y_0,\dot y_0) = \beta(z_0,\dot z_0)$ the equations behave identically (up to
a multiplicative constant). In this case, the angular momentum, 
$M^{01}=\frac{m}{2}(z \dot y
-y \dot z)$, is zero, and hence $y=\beta z$. The derivative of $M^{01}$ in 
the new variables is,
\begin{equation} \label{e13}
\frac{dM^{01}}{d \tau} = -\alpha(yz-1)M^{01}.
\end{equation}
For the system (\ref{e12a}), (\ref{e12b}), $M^{01}$ appears to converge,
in actual computations, to zero
independently of initial conditions. In fact, there are three typical 
behaviors of $M^{01}$ before it 
reaches zero: 
(a) An exponential growth (or oscillatory exponential 
growth) of $|M^{01}|$ followed by exponential decay (or oscillatory 
exponential decay) to zero, a scenario which occurs when 
$\dot M_0^{01} M_0^{01} > 0$. (b) An exponential decay (or oscillatory 
exponential decay) of $|M^{01}|$ to zero, which occurs when 
$M^{01} \ne 0$ and $\dot M_0^{01} M_0^{01} \le 0$. (c) $M^{01}$ is zero 
initially and remains zero. 

In fig. \ref{fig1} we present
the behaviors of cases (a) and (b). The numerical calculation of eqs. 
(\ref{e12a}), (\ref{e12b}),
were performed using the adaptive time step fifth-order Cash-Karp 
Runge-Kutta method \cite{NR}, and by fixed time step forth-order Runge-Kutta
method \cite{NR}. We choose $m=1$ for the mass and 
$\omega = 2.466$, $\alpha = 1$, $k=1$ and $f=10$. The initial conditions
used in fig. \ref{fig1}a were $y_0=1$, and $\dot y_0 = z_0 = \dot z_0 = -1$
($M_0^{01}=1$ and $\dot M_0^{01}=2$), and it is easy to observe the initial
exponential growth and the succeeding oscillatory exponential decay to zero.
In fig. \ref{fig1}b a typical behavior of case (b) is shown. The initial 
conditions are $y_0=3$, $\dot y_0 = 2$, $z_0 = \frac{1}{2}$ and $\dot z_0 = 0$
($M_0^{01}=\frac{1}{2}$ and $\dot M_0^{01}=\frac{1}{4}$), and the typical 
behavior of an exponential oscillatory decay to zero is shown. If fig. 
\ref{fig2}a and \ref{fig2}b we show the $(y,z)$ plane of fig. \ref{fig1}a 
and \ref{fig1}b which the solutions converge to a linear curve after a short 
time and then continue with oscillations on this line. Hence we also see in 
the phenomenon of entrainment in the van der Pol case.

After the angular momentum reaches zero there is no reason to investigate 
the two coupled eqs. (\ref{e12a}), (\ref{e12b}), which actually become 
uncoupled since the variables are proportional. Thus, one is left with the 
equation (we henceforth replace $\tau$ by the familiar designation $t$ 
appropriate to the nonrelativistic case):
\begin{equation} \label{e14}
\ddot x + \alpha (x^2-1) \dot x +kx = fx\cos (\omega t).
\end{equation}

A class of equations including eq. (\ref{e14}) were studied by Schmidt and 
Tondl \cite{Schmidt86} for the case for which both damping and nonlinear terms,
 as well as the homogeneous driving term are taken as small (as can be seen by
 scaling the dependent variable); they used perturbative techniques to study 
the instability. Cicogna \& 
Fronzoni \cite{Cicogna93} have studied the modification of chaos of the 
Duffing oscillator using an additional forcing term which can be proportional 
to $x$; a similar structure occurs in ref. \cite{Horwitz91}, as mentioned 
above. Such a forcing term occurs in the Mathieu equation (as a time-dependent
frequency) \cite{Glendinning}. Our study is not restricted to small
 coefficients; it utilizes numerical methods of wider applicability.

 In the following we will show the
chaotic properties of eq. (\ref{e14}), including the complex shape of the 
trajectory itself (which is reflected by a continuous frequency spectrum), 
and the fractal form of the Poincar\'e map, which in our case is just 
$(x(t_n), \dot x(t_n))$ where $t_n=t_0+nT$ ($T=\frac{2 \pi}{\omega}$ and $n$
is an integer number larger then $0$). The chaotic behavior is verified by 
the existence of a positive Lyapunov exponent. 

There exist several methods for calculating Lyapunov exponents from a set of 
first order differential equations \cite{{Geist90},{Wolf85}}. 
However, we preferred a new
method developed by Habib and Ryne \cite{Habib}, which is based on a 
technique using symplectic matrices. The symplectic computation of Lyapunov 
exponents is 
applicable whenever the linearized dynamics is Hamiltonian (which true for
our system). The basic advantage of the symplectic method is that it avoids
the renormalization and reorthogonalization procedures necessary in the usual 
techniques, and thus, leads to fast and accurate results (the method avoids an
additional error which can be caused by the renormalization and 
reorthogonalization procedures). 

According to the method of Habib and Ryne one linearizes the system, as a first
step in the computation of Lyapunov exponents. In our case, the linearization 
of eq. (\ref{e14}) is
\begin{equation} \label{e15}
\ddot \delta +\alpha(x_0^2-1) \dot \delta
+ (2 \alpha x_0 \dot x_0 +k -f \cos \omega t) \delta = 0,
\end{equation}
where $\delta=x-x_0$ and $x_0$ is the fiducial trajectory \cite{Wolf85}. 
Following Habib 
and Ryne, we change the variable to $\Delta=\delta \exp(-g(t))$, where 
$\dot g(t)=-\frac{\alpha}{2}(x_0^2-1)$, so that eq. (\ref{e15}) becomes,
\begin{equation} \label{e16}
\ddot \Delta - (\ddot g(t) + \dot g(t)^2 - k + f \cos \omega t) \Delta = 0.
\end{equation}
In order to find the Lyapunov exponents it is necessary to solve two
first order differential equations, \cite{Habib}
\begin{eqnarray} 
\frac{d\mu}{dt} &=& {1 \over 2} (s_{22}-s_{11})\cos a \nonumber \\
\frac{da}{dt} &=& s_{11}+s_{22}-(s_{22}-s_{11})\sin a \coth \mu \label{e17}.
\end{eqnarray}
The Lyapunov exponents are then $\pm \mu / t$, where $t$ is sufficiently large.
These equations are obtained as follows. In the general case of a Hamiltonian 
of quadratic form,
\begin{equation} \label{e18}
H({\bf Z},t) = {1 \over 2} \sum_{i,j=1}^{2m} s_{ij}Z_iZ_j,
\end{equation}
where ${\bf Z}=(q_1,q_2,\dots,q_m,p_1,p_2,\dots,p_m)$, the canonically 
conjugate coordinates and momenta, the equation of motion for 
${\bf Z}(t)=M(t){\bf Z}(0)$ implies that 
\begin{equation} \label{e18a}
\frac{d}{dt} M \tilde M = JSM \tilde M -M \tilde M SJ,
\end{equation}
where
\begin{equation} \label{e18b}
J = \pmatrix{
0 &{\bf 1} \cr
-{\bf 1} &0\cr}
\end{equation}
and $S=(s_{ij})$. The equations (\ref{e17}) are then obtained for a system 
with one degree of freedom (in the case that there is no term of the type $qp$
in the Hamiltonian), by recognizing that the most general form of the 
symplectic matrix $M$ is given by 
\begin{equation} \label {e18c}
M = e^{\mu(B_2 \cos a + B_3 \sin a)} e^{b B_1},
\end{equation}
where $B_1=i\sigma_1$, $B_2=\sigma_2$ and $B_3=\sigma_3$, and $\{\sigma_i\}$
are the Pauli matrices.

In the case of eq. (\ref{e19}), the $s_{ij}$ are : 
\begin{eqnarray} \label{e19}
s_{11} &=& -(\ddot g(t)+\dot g(t)^2 -k +f \cos \omega t) \nonumber \\
&=& \alpha x_0 \dot x_0 - \frac{\alpha^2}{4}(x_0^2-1)^2 +k-f\cos \omega t,\\
s_{22}&=&1. \nonumber
\end{eqnarray}
After finding the Lyapunov exponents, 
$\lambda_\pm = \pm\lim_{t \to \infty}{\mu (t) / t}$, 
of eq. (\ref{e16}) using eqs. (\ref{e17}), one must find the Lyapunov exponents
of the original system, eq. (\ref{e15}). By recognizing that asymptotically 
eq. (\ref{e16}) is $\Delta_{\pm} = \delta_{\pm}\exp(-g(t))=
\exp{\lambda_\pm t}$, one can then find the Lyapunov 
exponents of eq. (\ref{e15}) \cite{Habib}:
\begin{equation} \label{e20}
\chi_\pm = \lim_{t \to \infty} {1 \over t} (g(t) \pm \mu (t)).
\end{equation}

It is possible to summarize the Lyapunov exponents calculation of our
system (eq. (\ref{e14})) by five first ordered differential equations:
\begin{eqnarray} \label {e21}
\frac{d x_0}{dt} &=& p_0, \nonumber \\
\frac{d p_0}{dt} &=& -\alpha(x_0^2-1)p_0 -kx_0 +fx_0 \cos{\omega t},\nonumber\\
\frac{d g}{dt} &=& -\frac{\alpha}{2} (x_0^2-1), \\
\frac{d \mu}{dt} &=& {1 \over 2} (1-\alpha x_0p_0 + 
	\frac{\alpha^2}{4}(x_0^2-1)^2-k+f \cos{\omega t}) \cos{a}, \nonumber \\
\frac{da}{dt} &=& 1+\alpha x_0p_0 -\frac{\alpha^2}{4}(x_0^2-1)^2 +k -f 
	\cos{\omega t} -\nonumber \\
	&& (1-\alpha x_0p_0 + \frac{\alpha^2}{4}(x_0^2-1)^2-k+
	f \cos{\omega t}) \sin{a} \coth{\mu}, \nonumber
\end{eqnarray}
and the Lyapunov exponents can be found by eq. (\ref{e20}). The first two 
equations
calculate the fiducial trajectory. The third relation is needed for 
the calculation of the Lyapunov
exponents. The last two equations are just the explicit form of eqs. 
(\ref{e17}).

In order to locate some parameter values which lead to chaotic behavior, one 
must map the positive Lyapunov exponent in parameter space. Since our 
parameter space is represented in four 
dimensions \footnote{The parameter space is actually 
three dimensional since one can avoid the parameter $k$ in eq. (\ref{e14})
by introducing a 
rescaled time $s=\sqrt{|k|}t$.}, the mapping of chaotic regions are quite
difficult and it is necessary to focus on one or two parameters; we choose 
$\alpha$ to be the varying parameter since it controls the nonlinearity of 
the system.
\begin{figure}
\psfig{figure=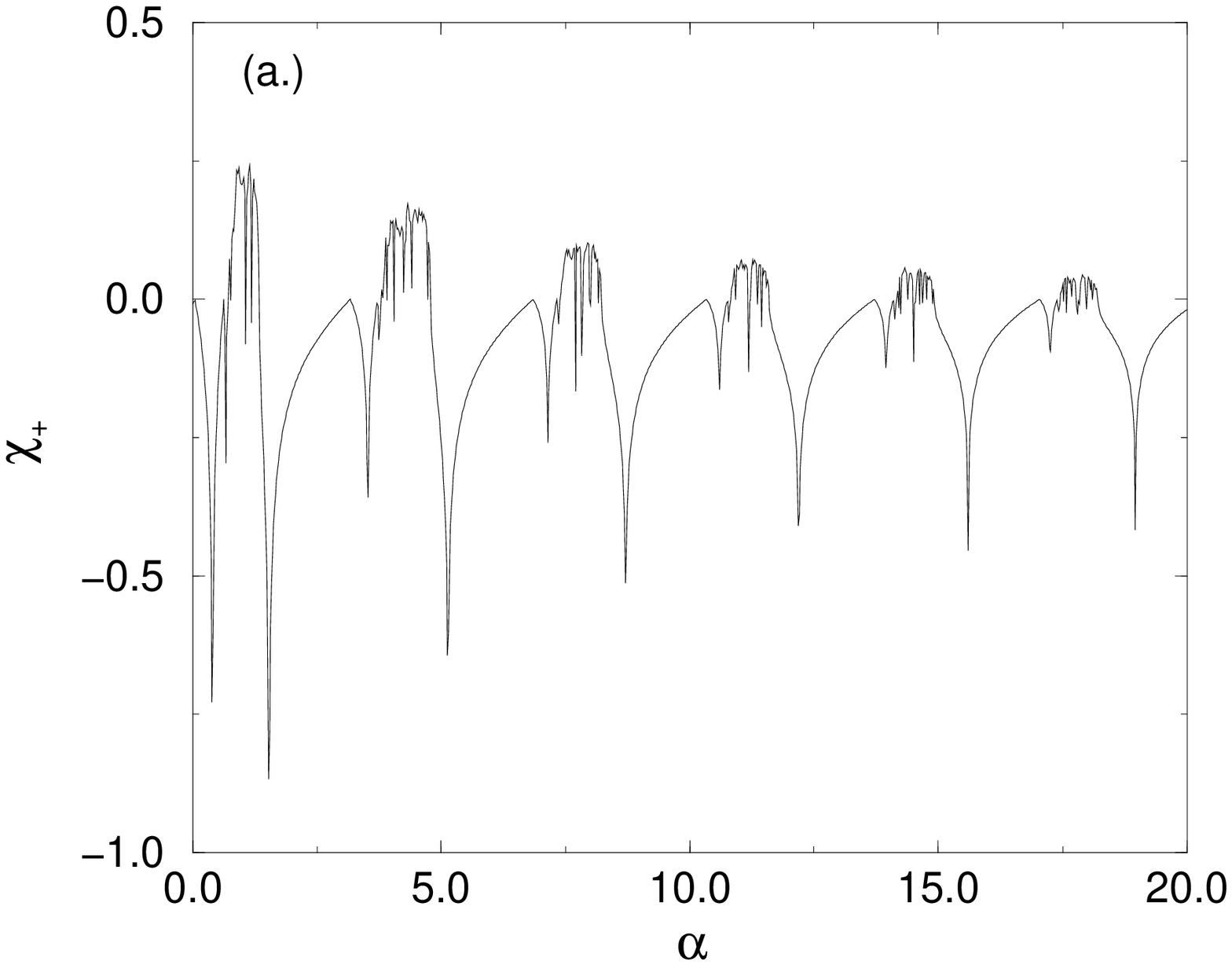,height=8cm,width=8.5cm}
\psfig{figure=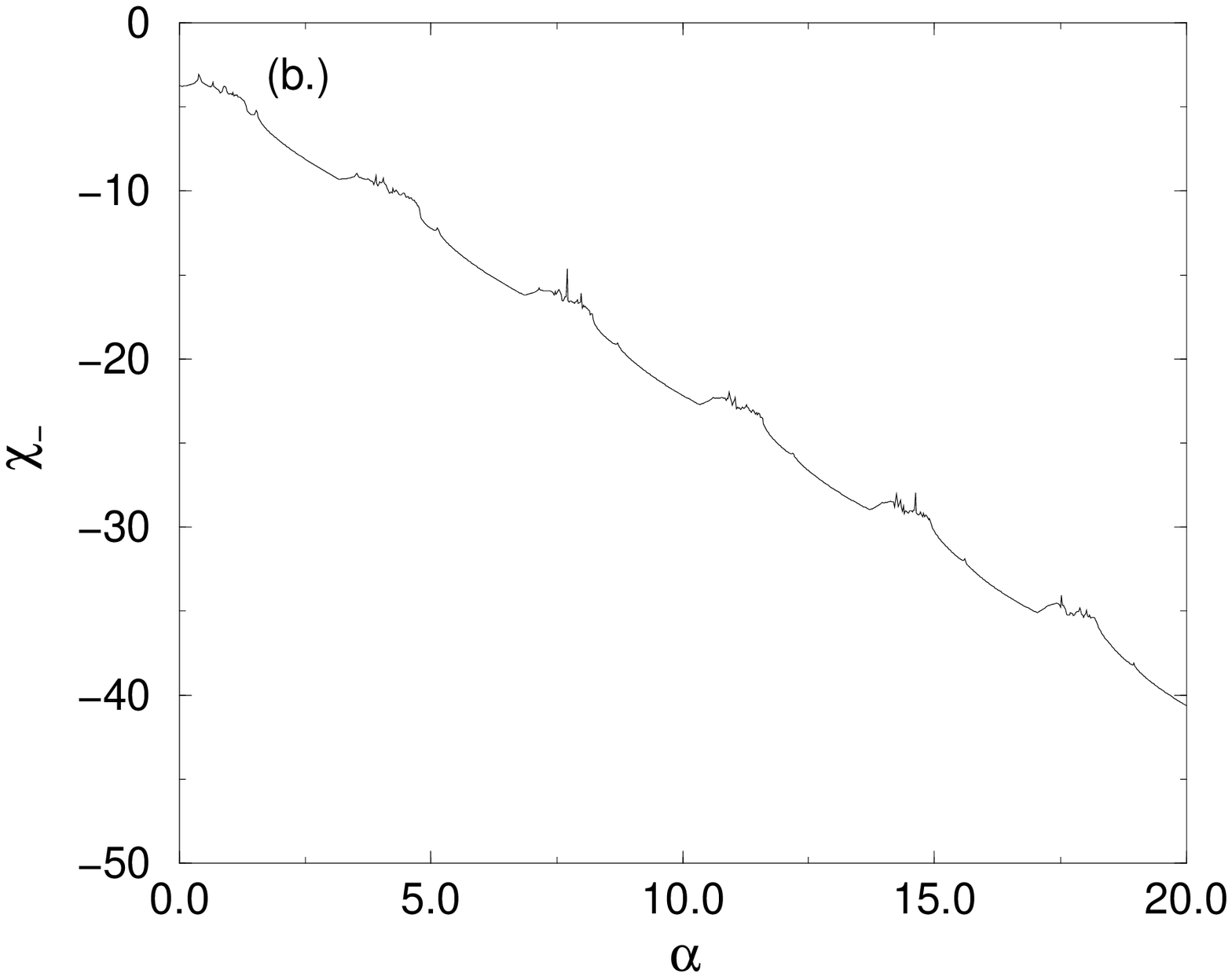,height=8cm,width=8.5cm}
\caption[tbp]{\label{fig3}
Lyapunov exponents as a function of $\alpha$ for a) positive Lyapunov exponent,
b) negative Lyapunov exponent.}
\end{figure}

In fig. \ref{fig3} we show Lyapunov exponents as a function of 
$\alpha$ (the other 
parameter values are $\omega=2.466$, $k=1$, $f=5$). We draw the Lyapunov 
exponents after $200000$ time units. The larger Lyapunov exponent is shown in 
fig. \ref{fig3}a, and it is 
easy to see that it becomes positive in a small region of the range of 
$\alpha$.
One can also identify some periodicity in $\chi_+(\alpha)$ multiplied by a 
decaying envelope. This ``periodicity'' gives a clue about some intrinsic 
frequency which
resonates with the ``external'' frequency, $\omega$. However, it seems that 
the gap between the chaotic regions is not exactly constant. A negative 
Lyapunov 
exponent is shown in fig. \ref{fig3}b. As expected, the absolute value of 
the negative Lyapunov exponent is much larger than the positive 
one. The negative Lyapunov exponent decreases approximately linearly with 
$\alpha$ (not including some regions parallel to chaotic regions with positive 
Lyapunov exponent) since $\alpha$ controls the dissipation of the system.

One of the fundamental questions in the numerical solution of chaotic systems 
is the validity of the numerical results. Since in every numerical integration 
there is an accumulative error, the integrated trajectory separates from the 
real trajectory in a very short distance. According to chaos theory, 
a nearby trajectory can generate a completely different trajectory (nearby 
trajectories actually separate 
exponentially), thus, one cannot guarantee the accuracy of the 
numerical calculation. The numerical trajectory is method dependent, and more 
than this, is accuracy dependent (if one changes the integration accuracy the 
trajectory behaves differently). One should expect, moreover, sensitivity 
to the integration method (or, equivalently, sensitivity to accuracy of the 
integration) if the system is chaotic, a fact which can used as an 
additional sign of chaotic behavior. Thus, it is clear that one cannot rely 
on the numerical results. The question is whether one can rely on the global
characteristic values, such as, Lyapunov exponents. It seems that the answer
is that one can indeed trust the global values, although the local behavior is
not assured \footnote{Greene \cite{Greene78} pointed out that there are huge 
numerical errors in calculating orbits in stochastic regions, yet these 
errors do not seem to change the fundamental character of the orbit. In his 
appendix he shows analytically that the error propagates anisotropically 
according to the direction of maximal instability. The stability transverse 
to these region was astonishingly well preserved and thus the boundaries of 
the region were visually stable regardless of instability along the orbit. 
Our results are in accordance with Greene's observations.}. 
It seems that although the trajectories behave differently
they appear to ``move'' on the attractor; they have the same global 
characteristic exponents. The influence of this sensitivity can be seen in fig.
\ref{fig3}a. One recognizes easily that there is a
clear smooth behavior of $\chi_+(\alpha)$ 
when it is less then zero (the system 
is not chaotic and nearby trajectories do not separate exponentially); once the
exponent becomes positive $\chi_+(\alpha)$ becomes less smooth. 
\begin{figure}
\psfig{figure=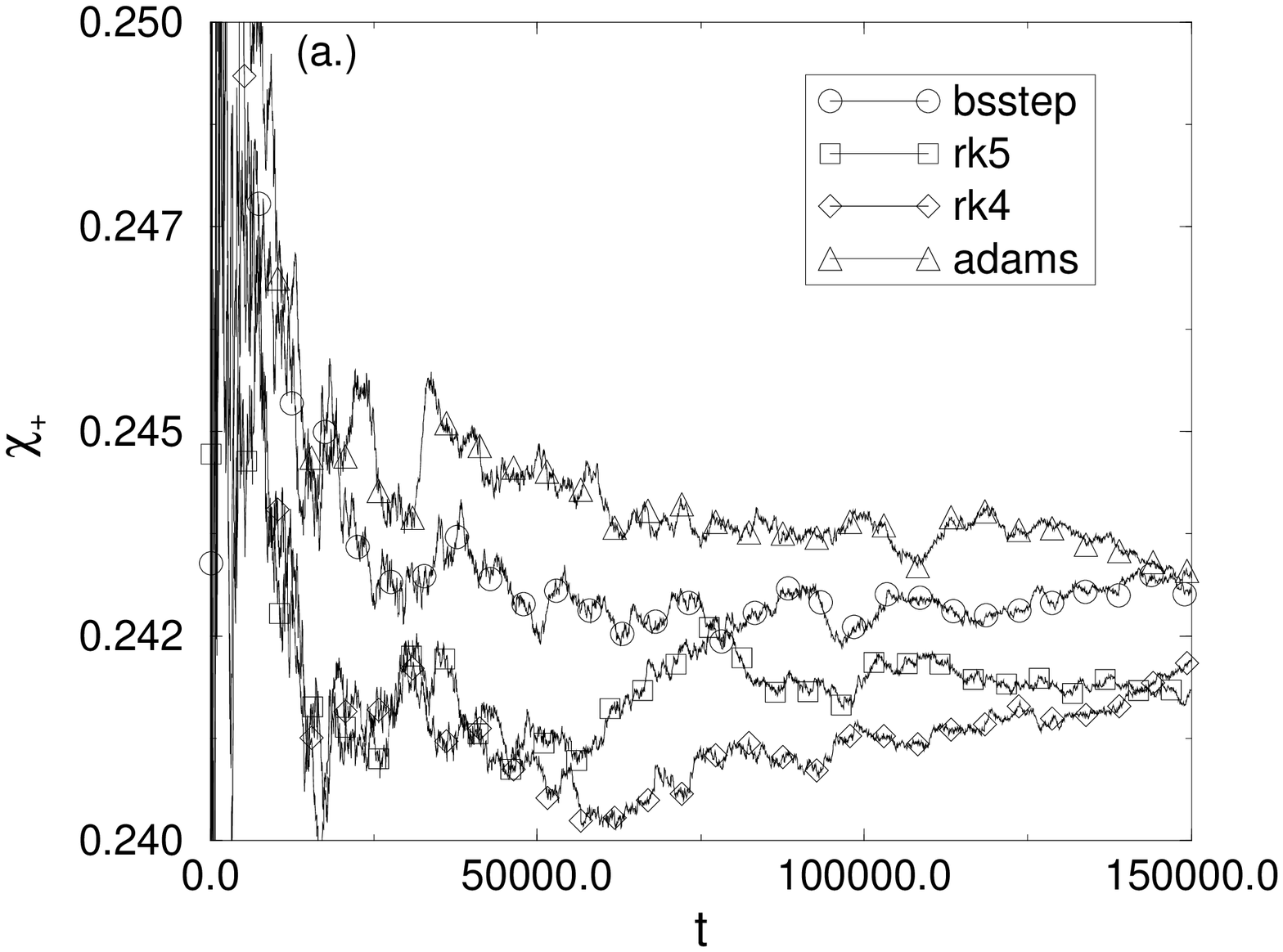,height=8cm,width=8.5cm}
\psfig{figure=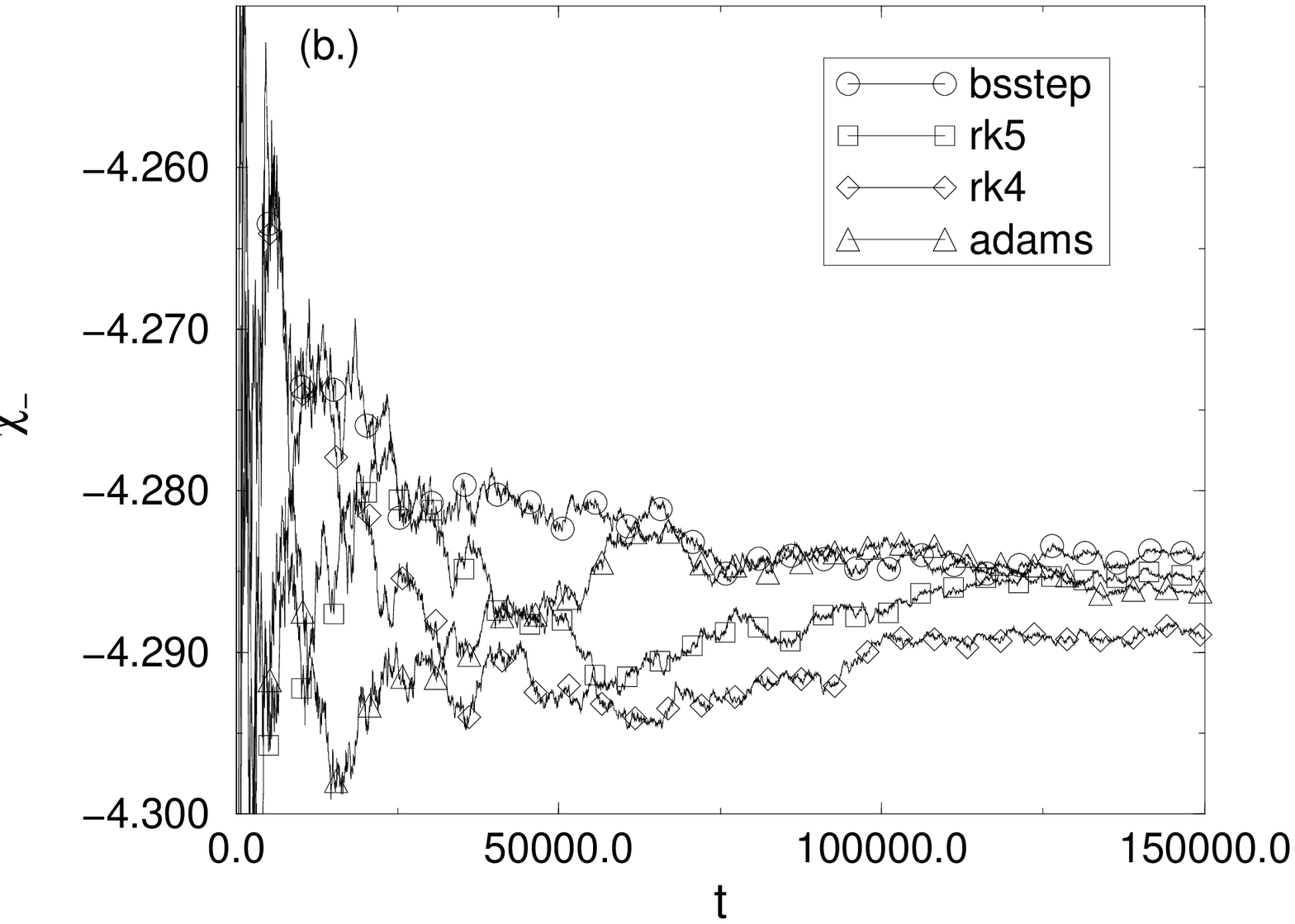,height=8cm,width=8.5cm}
\caption[tbp]{\label{fig4}
Convergence of Lyapunov exponents to a constant value, as a function of $t$, 
using four different integration methods for a) positive Lyapunov exponent,
b) negative Lyapunov exponent.
}
\end{figure}

\begin{figure}[t]
\psfig{figure=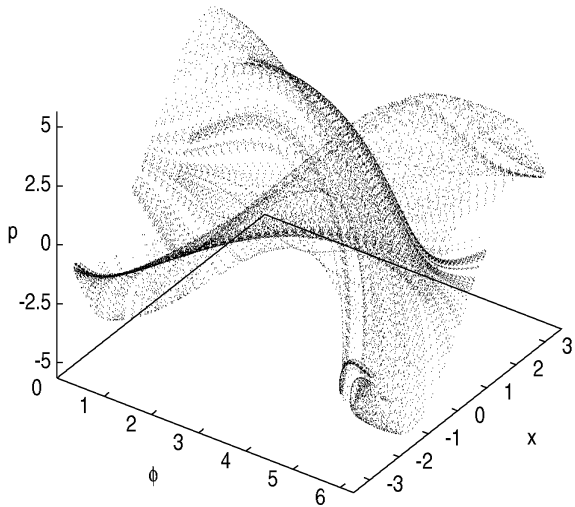,height=8cm,width=8.5cm}
\caption[tbp]{\label{fig5}
The attractor of relaxed relativistic van der Pol oscillator.}
\end{figure}
In order to verify
the claim that the Lyapunov exponents are the same for different integration 
techniques, we compare the results between four different methods: 1. 
Bulirsch-Stoer method using Richardson extrapolation \cite{NR}
(denoted by bsstep); 2. adaptive stepsize fifth Runge-Kutta method using 
Cash-Karp parameters \cite{NR} (denoted by rk5); 3. fourth Runge-Kutta method 
\cite{NR} (denoted by rk4); and 4. Adams integration method \cite{Nag}. 
We choose $\alpha=1.149$
since it gives the largest positive Lyapunov exponent as seen in fig. 
\ref{fig3}a.
The Lyapunov exponents are shown in fig. \ref{fig4} and it is clear that 
all methods lead to approximately the same exponents; $\chi_+=\approx 0.242$,
$\chi_-=\approx -4.286$. 

One can find other signs of chaotic behavior with the parameters used in fig.
\ref{fig3}
($\omega=2.466$, $k=1$ and $f=5$), and taking $\alpha=1.149$. 
In fig. \ref{fig5} the real phase space, $(\phi,x,p)$, where $\phi=\omega t$ 
mod $2 \pi$, is shown. It can be seen
clearly that the trajectory is very complicated and does not return to itself.
The phase space lies on a surface with some width; a fact which suggests a
noninteger fractal dimension. Using the Kaplan and Yorke conjecture
\cite{Kaplan79}, the fractal dimension would be :
\begin{equation} \label{e22}
D_L=j+{\sum_{i=1}^j\lambda_i \over |\lambda_{j+1}|} \approx 2+
{0.242 \over 4.286} \approx 2.06
\end{equation}
where $j$ is defined by the condition that $\sum_{i=1}^{j} \lambda_i > 0$
and $\sum_{i=1}^{j+1} \lambda_i < 0$ (the Lyapunov exponent in the time 
direction is, of course, zero).

\begin{figure}
\psfig{figure=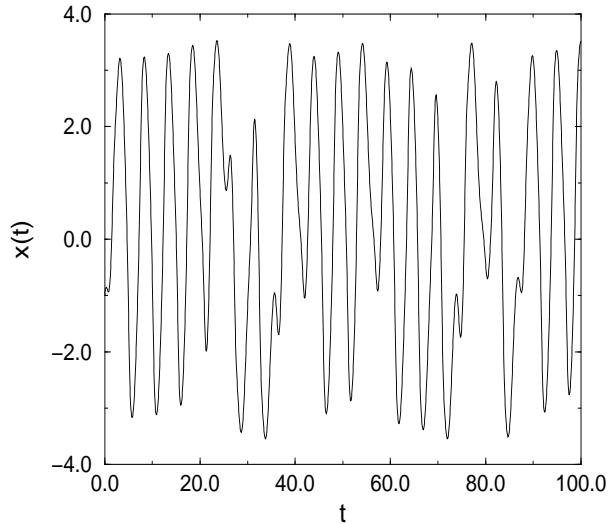,height=8cm,width=8.5cm}
\caption[tbp]{\label{fig6}
The complex behavior of $x(t)$ (we have used the notation $t$ instead of 
$\tau$ to describe the decoupled system).
}
\end{figure}
\begin{figure}
\psfig{figure=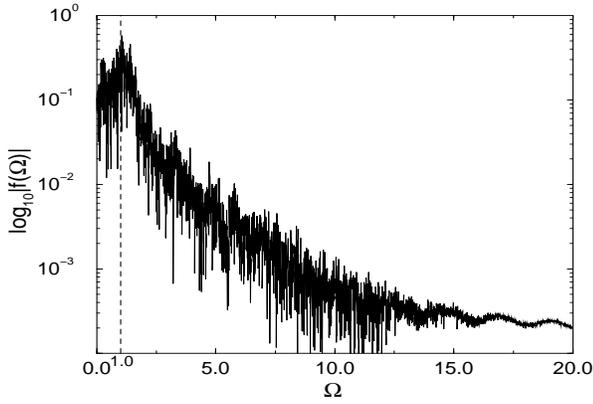,height=6cm,width=8.5cm}
\caption[tbp]
{
\label{fig7}
The power spectrum of fig. \ref{fig6}. The dashed line indicates the free 
oscillator frequency, i.e., $\Omega=\sqrt{k}=1$.
}
\end{figure}

No clear periodicity is seen in $x(t)$ , as shown in fig. \ref{fig6}.
The Fourier transform of $x(t)$ (fig. \ref{fig7}) shows a continuous 
frequency spectrum, which is a sign of chaotic behavior. In addition, there is
an approximate linear decay of the logarithm of the power spectrum, a fact 
which is connected to the exponential decay of the autocorrelation function 
\cite{Tabor89} (we also checked
the autocorrelation itself and its envelope indeed decays exponentially).
One sees a peak in the neighborhood of $1$, which is the frequency of the
undamped and unforced oscillator. One would expect resonant behavior in the 
neighborhood of half the driving frequency ($\Omega={\omega \over 2}$) on the 
basis of a comparison with the Mathieu equation \cite{Glendinning}. In our 
case, there is a apparent resonant region around this frequency $\Omega \sim
1.233$.

\begin{figure}
\psfig{figure=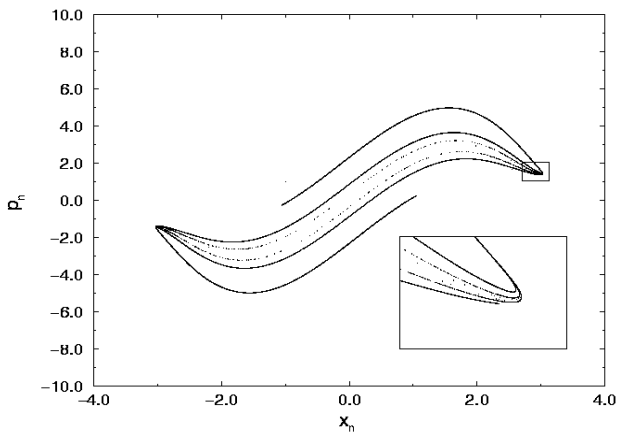,height=6cm,width=8.5cm,angle=-90}
\caption[tbp]{\label{fig8}
The Poincar\'e map of the relaxed relativistic van der Pol equation. The inset
shows an enlargement of the map.}
\end{figure}

Despite the difficulties mentioned above, we succeeded to draw a Poincar\'e
map for eq. (\ref{e14}). The map in this case is just the points on the plane
$\phi=0$, and can be written as $(x(t_n),p(t_n))$, where 
$t_n={2\pi \over \omega}n$ and $n$ is a nonnegative integer. The map is drawn
using two different methods : rk4, and the Adams method; one obtains the same 
mapping 
structure. This is a surprising result since, as mentioned before, one can 
not be sure of the accuracy of the trajectory $x(t)$ due to accumulation error 
which leads to a different 
trajectory behavior. The map shown in fig. \ref{fig8} shows smooth 
curves with stretching and folding phenomenon (more fine structure is seen 
in the inset). The map is antisymmetric, and obviously has a fractal dimension 
$\approx 1.06$ (following eq. (\ref{e22}) and ignoring the time direction). 

The entrainment phenomenon observed between space and time chaotic modes, both 
for the Duffing \cite{Horwitz91} and van der Pol case, for which $x$ and $t$
appear to approach exponentially to a smooth linear relation, even though both
$x$ and $t$ remain chaotic, presents a mechanism for the rapid convergence of 
damped relativistic systems to their classical limit. Underlying this smooth
behavior of $x(t)$, one might expect to see a characteristic property of the 
radiation field reflecting the chaotic behavior of both $x$ and $t$, in such 
systems.

We further remark that the stability of the Poincar\'e plots of fig. \ref{fig8}
indicates that computational deviations of $x$ are compensated to reach a
smooth curve by deviations in $t$. This stability is analogous to the 
stability of the calculation of the Lyapunov coefficients along a fiducial
curve that is calculated with an inevitable error. Both the Poincar\'e plots
and Lyapunov calculations appear to reflect properties of the attractor rather
that the local computed orbit.

We would like to thank W.C. Schieve for suggesting this problem and 
discussions in the initial stages of this work.

\end{document}